# Evolution of the density of states at the Fermi level of $Bi_{2-y}Pb_ySr_{2-x}La_xCuO_{6+\delta}$ and $Bi_2Sr_{2-x}La_xCuO_{6+\delta}$ cuprates with hole doping

M. Schneider[a], R.-S. Unger[a], R. Mitdank[a], R. Müller[b], A. Krapf[a], S. Rogaschewski[a],
H. Dwelk[a], C. Janowitz[a], R. Manzke[a]

[a] Humbold Universität, Insitut für Physik, Newtonstrasse 15, 12489 Berlin, Germany
[b] Physikalisch-Technische Bundesanstalt, Abbestrasse 2-12, 10587 Berlin, Germany

$Bi_2Sr_{2-x}La_xCuO_{6+\delta}$ and $Bi_{2-y}Pb_ySr_{2-x}La_xCuO_{6+\delta}$ high-$T_c$ superconductors in a wide doping range from overdoped to heavily underdoped were studied by x-ray absorption and photoemission spectroscopy. The hole concentration p was determined by an analysis of the Cu $L_3$-absorption edge. Besides the occupied density of states derived from photoemission, the unoccupied density of states was determined from the prepeak of the O K-absorption edge. Both, the occupied as well as the unoccupied density of states reveal the same dependence on hole doping, i.e. a continuous increase with increasing doping in the hole underdoped region and a constant density in the hole overdoped region. By comparing these results of single-layer BSLCO with previous results on single-layer LSCO it could be argued that besides the localized holes on Cu sites the $CuO_2$-planes consist of two types of doped holes, from which the so-called mobile holes determine the intensity of the prepeak of the O 1s absorption edge

## I. Introduction

In the last years, there was a continuing effort in the investigation of high temperature superconductors by photoemission, especially of the $Bi_2Sr_2CaCu_2O_{8+\delta}$ (BSCCO) system. One main topic was the evolution of the Fermi surface by doping [1]. While this momentum resolved information attainable with the resolving power of new analysers enabled a lot of insight into details of the electronic structure, a class of models like t-J-, Hubbard-model, and others yield results without explicit momentum resolution. Here the evolution of the spectral density at $E_F$, which in this paper is used synonymously with density of states, is treated without explicit k-resolution. While numerous studies exist, which cover only a limited doping range and enable a comparison to the above quoted theories just in this restricted range, we found just two publications covering the entire doping range [2, 3]. The focus of these studies on the $La_{2-x}Sr_xCuO_4$ (LSCO)- system was the underdoped range, where effects due to the pseudogap and the spectral weight renormalization were discussed for the occupied DOS. Interestingly in the overdoped range the DOS at $E_F$ was found to converge to a constant value and a similar trend was observed by Romberg et al. [4] for the unfilled DOS by EELS. Although the authors did not explicitly comment on this, in a naive view one would expect an increasing DOS with increasing hole concentration. Since the optimum doping regime with highest $T_c$ can be approached by doping from both sides, we found a strong demand for a comparative study on a different compound covering for the first time the occupied and the unoccupied DOS, to see whether the evolution of the DOS follows a common trend. One of the main research topics for HTC's is to find such universal properties like the pseudogap behaviour or, more general, the basic physics underlying the phase diagram.

For a systematic study of the doping dependence single-layer $Bi_2Sr_{2-x}La_xCuO_{6+\delta}$ (BSLCO) is one of the few other appropriate systems. It is the first member (n = 1) of the Bi-based cuprate family $Bi_2Sr_2Ca_{n-1}Cu_nO_{2n+4+\delta}$ with n = 1, 2, 3 (n - number of $CuO_2$-layers per unit cell). It's properties, especially the low $T_c$ and wide hole doping range when substituting Sr by La make it an ideal compound for a comprehensive study of the normal state at low temperatures. The Bi-cuprates possess a superstructure from reconstruction of the BiO-planes, which can be suppressed by partial substitution of Bi by Pb. For this single layer cuprate besides a suppression of the modulation structure [5] a $T_c$-enhancement is observed by the Pb-substitution [6]. For completeness, two series with and without additional Pb doping have therefore been investigated.

The determination of the number of holes in HTSC's is a contentious problem, but nevertheless crucial for a thorough discussion of the results. This problem is more complex for the La-doped Bi-2201 system since there is doping by replacement of Sr- and Bi-ions by La- and Pb-ions, respectively, and additionally hole doping by changes of the oxygen content in the material, all of which can affect the doping level of the $CuO_2$ sheets. Therefore, at least, methods for determining solely the oxygen content such as iodine titration or thermogravimetry are not sufficient for this system. These difficulties can in principle be overcome by measurements of the Hall coefficient respectively thermopower. Since $R_H$ is temperature dependent and must be normalized to equal volume and number of Cu-atoms, a quantitative calculation of $R_H$ is intricate. An example of the necessary procedures is given in [7]. The authors investigated different HTC's at optimum (highest $T_c$) hole concentration and found $R_H(T)$ to show similar values at room temperature when properly normalized. From this it was concluded, that a comparison of room temperature $R_H$ values of various HTC's at different doping levels to LASCO data at room temperature is the best choice. Furthermore and apart from the considerations of this study it can be stated, that the effective carrier concentration measured by the Hall effect must not be identical to the carrier concentration in the $CuO_2$-planes [8]. From these reasons we have chosen a more straightforward way by analysing the Cu $L_3$-absorption edge. This method has been used for LSCO [9, 10] and n = 2 [11-13] and n = 3 [14] Bi-based cuprates and is a direct measure of the carrier concentration in the $CuO_2$-planes.

In this paper we present the first data of the hole doping concentration obtained by analysis of the Cu $L_3$-absorption edge for the n = 1 Bi-based system. Based on our measurements the phase diagram of p versus $T_c$ for Bi-2201 can be displayed. It shows that the "universal" relationship or the so-called "bell shape" for the cuprates [15] holds not in the n = 1 Bi-based system. This is in agreement with the observations from Ando et al. [7] based on Hall measurements on the same material. Besides the occupied density of states at the Fermi level, extracted from photoemission (PES) data, an additional goal of this paper is the unoccupied density of states. For this, detailed investigations of the O K-edge have been performed. A comparison of the O 1s absorption edges of n = 1, 2, and 3 BSCO can be found in Müller et al. [16]. It is established that the states at the Fermi energy have primarily O 2p-character [17] and are connected with the prepeak feature of the O 1s $\rightarrow$ 2p absorption edge. The intensity of the prepeak is mainly interpreted as a direct measure of the hole concentration [4]. In contradiction to this assumption there are hints already in this previous experiment [4] which will be confirmed here by a large number of measured samples that for hole overdoping the intensity of the prepeak is not further increasing but saturating. Therefore an alternative explanation of the origin of the O 1s prepeak consistent with all observations has to be found.

## II. Samples

$Bi_2Sr_{2-x}La_xCuO_{6+\delta}$ and $Bi_{2-y}Pb_ySr_{2-x}La_xCuO_{6+\delta}$ are Bi-based single layer cuprates where the hole doping is controlled by replacing $Sr^{2+}$ with $La^{3+}$. With increasing La content the hole concentration decreases. Additional Pb doping (substituting $Bi^{3+}$) gives a suppression of the incommensurate modulation structure [5] and a moderate increase of the transition temperature [6].

All samples were sintered pellets prepared by standard powder metallurgical methods from mixtures of $Bi_2O_3$, $La_2O_3$, CuO, and SrCO with a composition chosen to give a stoichiometric weight of $Bi_2Sr_{2-x}La_xCuO_{6+\delta}$. The preparation was identical for all samples except for the final sintering, which was carried out at 890°C for samples with high La content and at 800 °C for samples without La. The higher sintering temperatures for samples with high La content are necessary in order to avoid phase separation. The sintering was carried out in air, and samples were cooled in the oven to room temperature after sintering. The Pb doped samples were additionally annealed in vacuum at 550°C for 5 h to increase the transition temperature [6].

The chemical composition and the transport properties of the samples were controlled by measuring the X-ray emission spectra (EDX) and ac susceptibility, respectively. The results are summarized in Table I. EDX measurements were performed on different sample positions and show the uniformity of the stoichiometry in the sample. The optimally doped samples show in the ac susceptibility a sharp transition $\Delta T$ of 2 K, as an indication for a single phase sample. This high quality is reached for the entire series. The existence of more than one superconducting phase would result in very broad transitions or different transition steps, which were not observed for optimally prepared samples [18]. An example for the ac susceptibility result is given in Fig 1 together with a resistivity measurement on the same sample. For the whole series of samples we found that the $T_{c, zero}$-value of the resistivity curve is equal within an accuracy of $\pm$ 0.5 K with the onset temperature value of the ac-susceptibility curve. A similar result has been observed for Bi-2201 single crystals by Ono and Ando [19] and by ourselves. In this respect the susceptibility value can be regarded as a conservative lower bound. The estimation of $T_{c, onset}$ in the resistivity measurement yields systematically higher transition temperatures compared to the susceptibility measurement.

The optimally doped sample is reached in the Pb-free series with x = 0.33 and $T_c$ = 18 K and in the series with additional Pb doping with x = 0.36 and $T_c$ = 34 K. Samples without La show $T_c$ = 6 K for the Pb-free series and $T_c$ = 12 K for the Pb-doped series. Superconductivity vanishes on the hole underdoped side at x = 0.78 and x = 0.55, respectively. The series $Bi_{2-y}Pb_ySr_{2-x}La_xCuO_{6+\delta}$ has a average Pb-concentration of y = 0.3.

## III. Experiment

The experiments were carried out by extensive investigations employing synchrotron radiation. The PES-measurements were performed at HASYLAB, Hamburg, by the use of linearly polarized radiation emitted from the high-resolution 3m normal-incidence monochromator HONORMI at beam line W3.2. For the measurements discussed here 18 eV photon energy was used. The energy distribution curves (EDC) were recorded with a hemispherical deflection analyzer with a total acceptance angle of 1° [20]. The measurements were performed at T = 50 K and with a total energy resolution of 55 meV as determined from an Au Fermi edge. The access to the density of states where k-resolution is not needed was ensured by the polycrystalline, ceramic samples.

**Table I** (a) Chemical compositions of the sample series $Bi_2Sr_{2-x}La_xCuO_{6+\delta}$ and (b) $Bi_{2-y}Pb_ySr_{2-x}La_xCuO_{6+\delta}$ obtained by EDX (Bi, Sr, La, Cu). The values are averages over different surface areas. Additionally the transition temperature $T_c$ taken from the onset temperature of the ac-susceptibility curve and the hole concentration as result of the Cu $L_3$-absorption edge are listed.

| (a) | Bi | Sr | La(x) | Cu | $T_c$ (K) | p |
|---|---|---|---|---|---|---|
| | 2.03 | 1.72 | 0 | 1.25 | 6 | 0.19 |
| | 1.95 | 1.92 | 0.10 | 1.03 | 8.5 | 0.22 |
| | 2.02 | 1.82 | 0.15 | 1.02 | 11 | 0.20 |
| | 1.93 | 1.85 | 0.23 | 0.99 | 15 | 0.19 |
| | 1.96 | 1.66 | 0.31 | 1.01 | 17 | 0.17 |
| | 1.99 | 1.66 | 0.33 | 1.01 | 18.5 | 0.17 |
| | 1.91 | 1.69 | 0.4 | 0.98 | 16 | 0.19 |
| | 1.97 | 1.56 | 0.45 | 1.02 | 13 | 0.15 |
| | 1.97 | 1.44 | 0.56 | 1.04 | 8 | 0.12 |
| | 1.98 | 1.21 | 0.78 | 1.03 | 0 | 0.07 |

| (b) | Bi | Pb(y) | Sr | La(x) | Cu | $T_c$ (K) | p |
|---|---|---|---|---|---|---|---|
| | 1.83 | 0.19 | 1.92 | 0 | 1.06 | 12 | 0.22 |
| | 1.69 | 0.30 | 1.92 | 0.16 | 1.08 | 24 | 0.19 |
| | 1.61 | 0.30 | 1.65 | 0.36 | 1.08 | 34 | 0.16 |
| | 1.66 | 0.34 | 1.52 | 0.40 | 1.10 | 21 | 0.12 |
| | 1.63 | 0.34 | 1.53 | 0.41 | 1.01 | 31 | 0.15 |
| | 1.60 | 0.43 | 1.39 | 0.50 | 1.09 | 5 | 0.12 |
| | 1.69 | 0.30 | 1.35 | 0.55 | 1.09 | 0 | 0.10 |
| | 1.69 | 0.30 | 1.31 | 0.58 | 1.12 | 0 | 0.11 |
| | 1.73 | 0.27 | 1.27 | 0.64 | 1.09 | 0 | 0.06 |

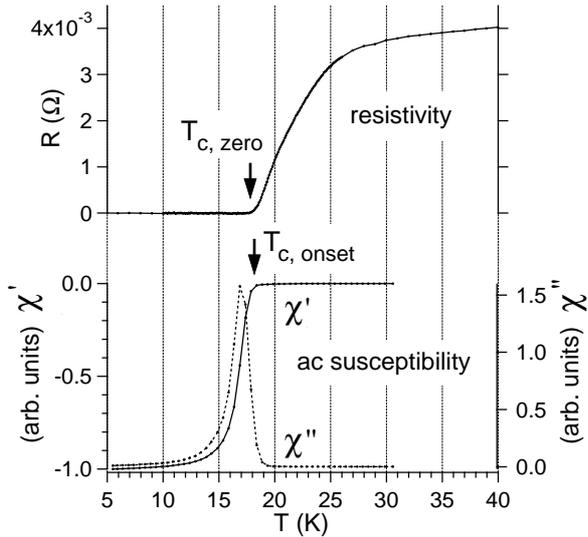

**Fig 1.** Resistivity and ac susceptibility of $Bi_2Sr_{2-x}La_xCuO_{6+\delta}$ at optimum doping (x = 0.33). Note that $T_{c,\,onset}$ of the ac susceptibility curve is at the same temperature as $T_{c,\,zero}$ of the resistivity curve.

The X-ray absorption experiments on the oxygen K-edge and copper L-edge were carried out at BESSY, Berlin, at different beamlines. With the experimental set-up it was possible to collect simultaneously both, the fluorescence signal by a windowless high purity Ge-detector and the total electron yield by a channeltron detector. The vacuum in the chamber was in the $10^{-10}$ mbar range during all measurements. In addition to that, the photoemission current of a clean gold mesh with 80% transmission was recorded for normalization of the spectra to the incoming beam intensity. The photons impinged parallel to the normal direction of the sample and were detected at an angle of 45° with respect to the sample normal.

The Cu $L_3$-edge X-ray absorption measurements were performed at liquid-nitrogen temperature at the U49/2-PGM-2 undulator beam line. The overall energy resolution was set to 350 meV. The O K-edge measurements were carried out for the Pb-free sample series at the VLS-PGM beam line at room temperature and a energy resolution of 400 meV, for the Pb-doped sample series at the U49/2-PGM-2 undulator beam line at liquid-nitrogen temperature with an overall energy resolution of 200 meV. The photon energies were calibrated with an accuracy of 0.1 eV by using the O K-edge absorption peak at 530.1 eV and the Cu $L_3$ white line at 932.1 eV of a CuO reference. The X-ray fluorescence-yield spectroscopy method is bulk-sensitive, the probing depth is 1000-5000 Å. For all measurements clean surfaces were obtained by in situ scraping with a diamond file.

## IV. Copper $L_3$-edge and hole concentration

Quantitative estimations for the hole concentration of $CuO_2$-plane were obtained from the Cu $L_3$-edge absorption spectra. A typical spectrum is shown in Fig 2. It includes a straight line for background subtraction and two pseudo-Voigt functions to fit the two absorption lines [12]. Besides the main line corresponding to the transition $2p^63d^9 \rightarrow 2p^53d^{10}$ an additional shoulder occurs from the oxygen ligand $2p^63d^9\underline{L} \rightarrow 2p^53d^{10}\underline{L}$, with $\underline{L}$ representing a hole on the oxygen 2p orbital. From the relative intensities of the two peaks the hole concentration can be determined by the intensity ratio $p = I(3d^{10}\underline{L}) / \{ I(3d^{10}) + I(3d^{10}\underline{L}) \}$. This method has been used to examine the hole concentration in the double-layer BISCO-system [11 – 13], after having been used before for LSCO [9, 10]. The background subtraction and fit has been performed in a manner reported by Ghigna et al. [12]. From this procedure the error in the density of doped holes is estimated to be lower than 5%. Karppinen et al. [13] have specified an error for this method of $\pm$ 0.02 holes.

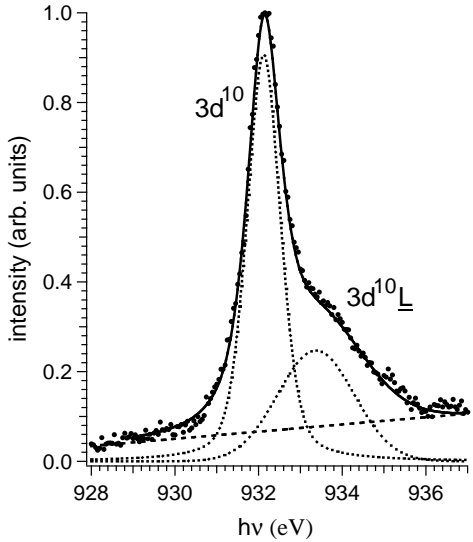

*Fig 2.* Cu-$L_3$ XAS-spectrum of a $Bi_2Sr_{1.67}La_{0.33}CuO_{6+\delta}$ sample. Dots are the experimental data, dashed lines refer to fits of the individual peaks by pseudo-Voigt profiles, full line gives the resulting convolution of the spectrum. The background was subtracted by a straight line, whose slope was chosen to match the high and low energy side of the spectrum. For details see the text.

The results of the hole concentrations for the Pb-free and Pb-doped single-layer BISCO series are given in Fig 3 (a). A proportionality between hole concentration and La content is clearly seen. From linear fits to the data the following linear relations between the hole concentration p and the Lanthanum concentration x can be approximated for $x < 0.8$: for the Pb-free series $p = (0.24 - 0.21\, x)$ and for the Pb-doped series $p = (0.23 - 0.22\, x)$. In Fig 3 (b) the relationship between the value of $T_c/T_{c,max}$ and the hole concentration p is shown, including the fit $T_c = T_{c,max}\left(1 - 250\left(p - p_{optimal}\right)^2\right)$. The dashed line shows the commonly employed empirical relation [15] derived from double-layer cuprate systems. In agreement with the single crystals studied by Ando et al. [7] and the ceramics studied partly in the first work by Sales and Chakoumakos [21], the single-layer BSLCO system shows a faster drop on both, the underdoped and overdoped sides of the phase diagram. Looking to the experimental points very closely one could realize a suppression of $T_c$ for 12.5% (1/8) hole concentration. A similar drop has also been observed on $n = 1$ BSLCO single crystals but the achievement of this small regime in the phase diagram is extremely critical and has to be investigated further.

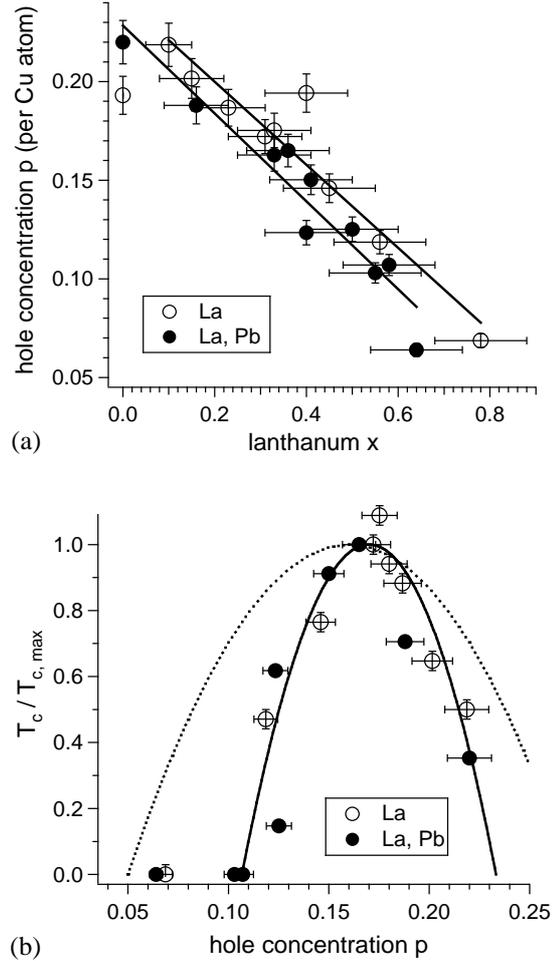

*Fig 3.* a) Hole concentration p of $Bi_2Sr_{2-x}La_xCuO_{6+\delta}$ and $Bi_{2-y}Pb_ySr_{2-x}La_xCuO_{6+\delta}$ as determined from the Cu-$L_3$ ligand absorption edge with respect to the La-content x. The lines are fits to the data points of the Pb-free and Pb-doped samples. b) Relationship between $T_c/T_{c,\,max}$ and the hole concentration p of $Bi_2Sr_{2-x}La_xCuO_{6+\delta}$ and $Bi_{2-y}Pb_ySr_{2-x}La_xCuO_{6+\delta}$. Parameters $T_{c,\,max} = 17$ K for the Pb-free series and $T_{c,\,max} = 34$ K for the Pb-doped series. The solid line corresponds to a parabolic fit of the experimental values (see text). The single layer systems $Bi_2Sr_{2-x}La_xCuO_{6+\delta}$ and $Bi_{2-y}Pb_ySr_{2-x}La_xCuO_{6+\delta}$ reveal with respect to the so-called universal curve (dashed line [15]) a faster drop on the underdoped and overdoped sides.

## V. Density of states at the Fermi level

In Fig 4 the photoemission (PES) spectra in the vicinity of the Fermi level are shown together with the X-

ray absorption (XAS) spectra of the O K-edge which represent the dominantly oxygen derived unoccupied states. The photoemission and absorption measurement were made consecutively on the same sample. The PES spectra have been normalized to the maximum intensity of the valence band at -3.5 eV binding energy, similar to [2].

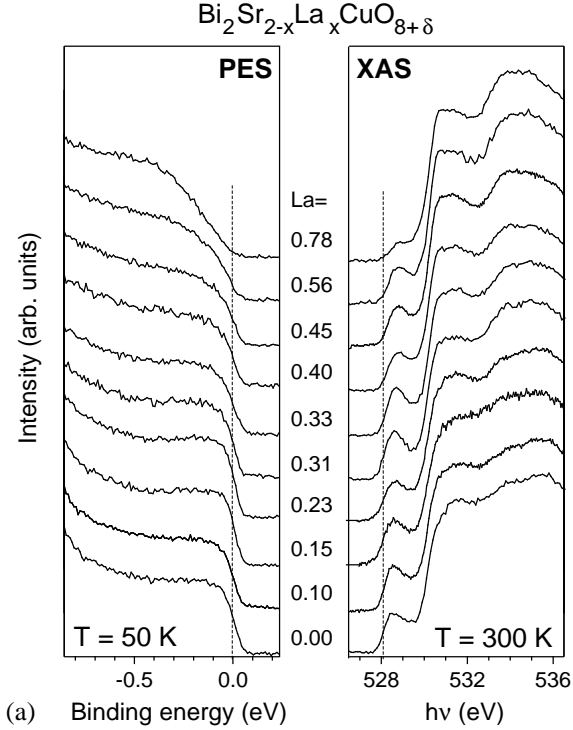

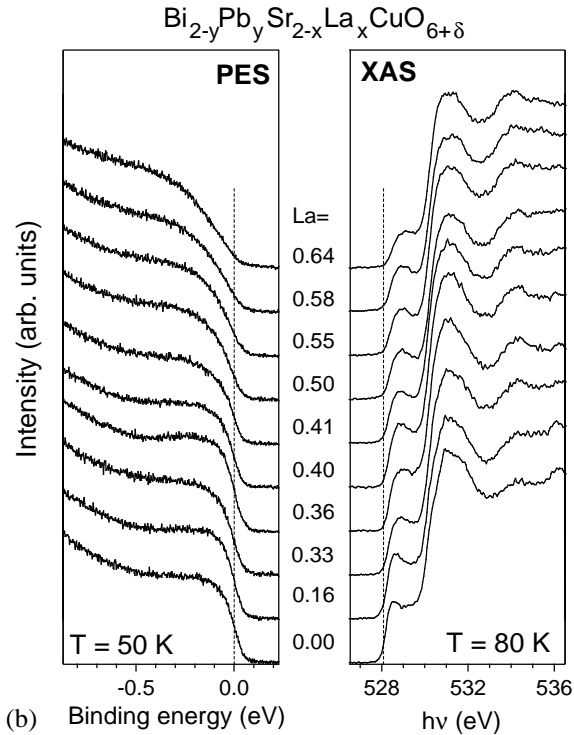

*Fig 4.* Photoemission spectra (PES) in the region near the Fermi energy (zero binding energy) and X-ray absorption spectra (XAS) of the O K-edge with varying La-content of $Bi_2Sr_{2-x}La_xCuO_{6+\delta}$ (a) and $Bi_{2-y}Pb_ySr_{2-x}La_xCuO_{6+\delta}$ (b).

The photoemission onset clearly reveals a Fermi-Dirac distribution for low La concentrations up to about x = 0.4 due to the metallic character of the samples. With increasing La concentrations, the intensity at the Fermi energy decreases. The metal-to-insulator transition appears between x = 0.45 and x = 0.56 for the lead-free samples and x = 0.55 and x = 0.58 for the Pb-doped samples, respectively. The spectra are comparable with the results for $La_{2-x}Sr_xCuO_4$ [2] unaffected of the instrumental resolution. In order obtain a measure of the relative density of states at the Fermi level the photoemission intensity on the occupied side of $E_F$ has been determined by integration of the spectral intensity in an energy window comparable with the energy resolution of 55 meV. In this case the error of the measured spectra is dominated by the error of the photon beam intensity $I_0$ used for the normalisation of the spectra. It can be estimated to about 3 %. The results are shown in Fig 5 (a) (upper picture) for the Pb-free and Pb-doped series.

The absorption measurements of the O K-edge performed for the lead-free samples $Bi_2Sr_{2-x}La_xCuO_{6+\delta}$ at room temperature and as result of the improvement of the experimental equipment for the Pb-doped samples $Bi_{2-y}Pb_ySr_{2-x}La_xCuO_{6+\delta}$ at liquid-nitrogen temperature are depicted in the right panels of Fig 4 (a) and (b), respectively. However, the XAS spectra are affected by the sample temperature only to a very small extent. All the absorption curves were normalized to the incident photon flux ($I_0$) and to the intensity at 600 eV photon energy. After this, a correction for self absorption effects has been carried out [22]. On the unoccupied side of $E_F$ the density of states is formed by the so-called prepeak of the O 1s absorption line. For the determination of the intensity of the prepeak, an extensive procedure was necessary and was done following the description of ref. [12] and [13]. In a first step, the absorption edge at 530 eV has been fitted by two Gaussian functions. In a second step, this fit was subtracted from the spectrum. Then the spectral intensity of the isolated prepeak was integrated in an energy region from 527 to 530 eV. The error of the spectra is dominated by the normalization to the incident photon flux and the self absorption correction process. It can be estimated to less than 6% [23, 24]. The results for both, the Pb-free series and the Pb-doped series is shown in Fig 5 (a) (lower picture) what corresponds qualitatively to the empty density of states at $E_F$.

We fitted the dependence of the DOS versus p as depicted in Fig 5 (a) on trial with one straight line and also with two straight lines intersecting at an arbitrary point i.e. having different slopes. We found significantly lower chi$^2$ values for the fit with two straight lines, one of them having a slope near zero. This corroborates our view that the DOS saturates at a certain doping level.

### VI. Discussion

With regard to the occupied density of states at the Fermi level $D(E_F)$ we found that our results on BSLCO are basically in agreement with previous observations on the LSCO-system [2, 3]. In Fig 5 the hole dependent $D(E_F)$ of single-layer cuprates is compared.

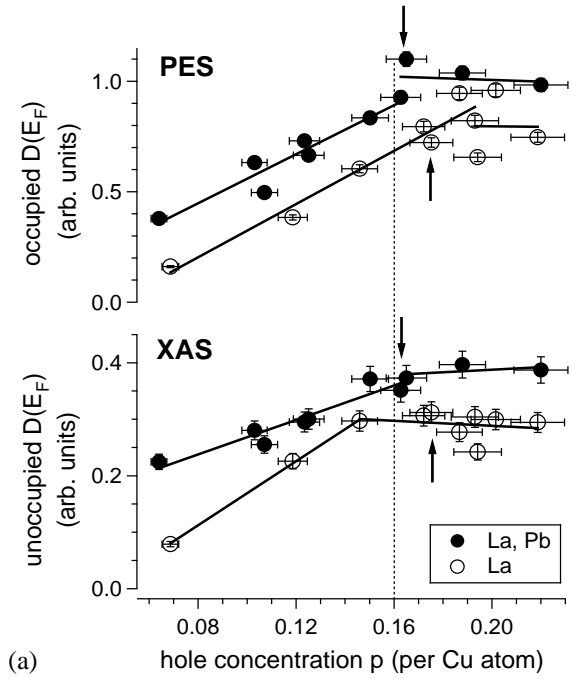

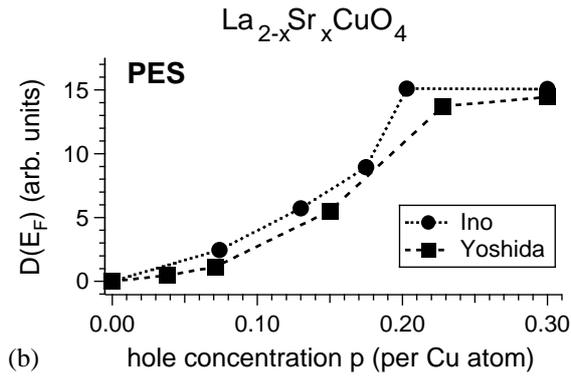

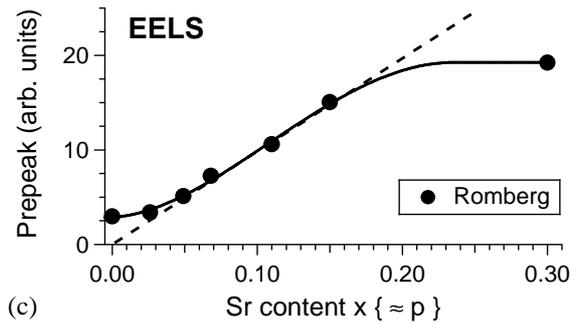

*Fig 5.* a) Occupied (from PES) and unoccupied (from XAS) density of states at the Fermi energy, $D(E_F)$, of $Bi_2Sr_{2-x}La_xCuO_{6+\delta}$ and $Bi_{2-y}Pb_ySr_{2-x}La_xCuO_{6+\delta}$ as a function of the hole concentration p. For better comparison p = 0.16 is marked by a dashed line. In addition, samples of optimally doping (highest $T_c$) are marked by an arrow. Lines are the result of fits. Previous work performed on $La_{2-x}Sr_xCuO_4$ is given in b) and c). The occupied density of states at the Fermi level $D(E_F)$ from PES of Ino et al. [2] and Yoshida et al. [3] is shown in (b) and of the prepeak intensity (unoccupied states) from electron energy-loss spectroscopy (EELS) of the O1s line from Romberg et al. [4] is given in (c).

In both cases the density of states increases with hole concentration in the underdoped regime until reaching a constant value for overdoped samples.

The difference between the two single-layer systems is the hole concentration where the change to the constant density of states occurs: For the LSCO-system the change occurs at p ≈ 0.20 [2, 3] and in the BSLCO-system at p ≈ 0.17, i.e. directly adjacent to optimum doping. While for LSCO the authors studied only two [2] or three [3] overdoped samples, for BSLCO we studied the overdoped region in great detail on a large number of samples. Therefore the evolution into constant density of states for the overdoped samples should now be viewed as confirmed and established.

It is interesting to relate the results of the occupied density of states to that from angular resolved photoemission. The change of $D(E_F)$ with increasing hole concentration can then be interpreted in relation to the evolution of the Fermi surface with doping studied in detail for LSCO [25, 26]. In lightly-doped cuprates the density of the states at the Fermi level is found to be formed by so-called "Fermi arcs" [26], which build up part of the hole-like Fermi surface around the nodal directions, i.e. (0,0) – (π,π) and equivalents, of the Brillouin zone. It should be mentioned that the Fermi arcs are due to a small sharp peak crossing $E_F$ in the nodal direction, but is observed until yet only in the second Brillouin zone [26]. In addition, these first time observations have to be confirmed in other single-layer cuprates, like e.g. n = 1 BLSCO. The spectral weight due to the arcs increases with hole doping in line with the $D(E_F)$ increase of the underdoped regime shown in Fig 5 in comparison. On the other hand, the flat band around the anti-nodal (0,0) – (π,0) direction of the Brillouin zone, located for low hole concentrations well below $E_F$, gradually moves up to $E_F$ with increasing hole doping for LSCO [2]. In addition for LSCO with stronger overdoping the energy gap around (π,0) closes and an electron-like Fermi surface is formed at p = 0.30. In principle, from such a closed electron-like Fermi surface constant density of states $D(E_F)$ could be expected, although a similar closure has not been observed in BISCO yet. For our polycrystalline samples a maximum hole concentration of p = 0.22 was obtained. It can be speculated, that for the single crystalline samples of BISCO used in angular resolved photoemission studies for Fermi surface mapping a similar limiting value exists, which is decisively below the value of p = 0.30 for LSCO, so that the LSCO is more heavily overdoped. Also different in n = 1 BSLCO is that the constant $D(E_F)$ begins already at about optimum doping of p = 0.16 and p = 0.17 for Pb-doped and Pb-free samples, respectively. For optimum hole doping, the flat band around the anti-nodal direction of the band structure is certainly below $E_F$ [27] and can therefore not be the explanation for a constant $D(E_F)$.

From the theoretical side a comparison to state of the art calculations taking correlation effects into account is worthwhile. One-band Hubbard-model calculations [27] with a next-nearest-neighbour hopping t' (t'/t = -0.35) predict similar behaviour as observed in the experiments.

Here the maximum intensity on the Fermi surface is reached at a doping level of p = 0.22, in agreement with LSCO [2, 3] but distinctly above the value of n = 1 BSLCO. At the doping level of p = 0.35 a decrease of the intensity is found in this theory. This doping level was not reached in our experiments.

The unoccupied density of states at the Fermi level, extracted from the prepeak of the O K-edge shows a very similar behavior as the occupied states including the convergence to a constant value in the hole overdoped region as described before (see Fig 5 (a) for n = 1 BSLCO). This can be seen as hint to the consistency of the interpretation of our data in form of density of states at the Fermi level derived from the filled and unfilled side of $E_F$. Previously, the prepeak of the O K-edge was interpreted in a straight forward manner as reflecting the hole concentration directly [4]. If this would be true a similar intensity behavior of the prepeak would be expected as the oxygen ligand line at the Cu $L_3$-edge, i.e. one would expect a continuous increase in the underdoped as well as in the overdoped regime. Clearly, the experimental results are contradictory to this and an alternative explanation of the origin of the O 1s prepeak has to be found.

Now the results of the LSCO single layer system from literature should be compared with that of BSLCO of the present work. Fig 5 (b) shows the density of states at the Fermi level derived from PES on $La_{2-x}Sr_xCuO_4$ by Ino et al. [2] and Yoshida et al. [3]. In agreement with BSLCO, in both experiments on LSCO $D(E_F)$ increases continuously for low doping and reveals a plateau for the overdoped region. In addition to this, also the unoccupied states derived by Romberg et al. [4] from electron energy-loss spectroscopic (EELS) investigations of the O 1s prepeak reveal a similar increase of the intensity and, in principle, also the plateau in line with our XAS data and the PES results discussed above, although they studied only one overdoped sample (see Fig 5 (c)). Romberg et al. [4] draw two lines to their data points, one solid line following the evolution of the data point by point and a second dashed straight line indicating a linear relationship between hole doping and prepeak intensity. According to the large number of data of the overdoped regime discussed here, for BSLCO as well as LSCO, we can now conclude that the linear increase is wrong and that the evolution into constant density of states for overdoped single layer cuprates is established. This result, on the other hand, needs a better interpretation for the origin of the O 1s prepeak than simply a count of holes in the $CuO_2$-plane. An explanation could be that besides the localized holes on Cu sites, upon doping two types of holes are introduced into the samples both located in the $CuO_2$-planes. Both types of doped holes determine the intensity of the oxygen ligand line at the Cu $L_3$-edge but only one type determines the intensity of the prepeak of the O K-edge. The O 1s prepeak has been called "mobile carrier peak" by Abbamonte et al. [29]. Thus we would suggest that it are the "mobile holes" which are measured by the O 1s prepeak intensity and which are responsible for the superconductivity of the $CuO_2$-planes.

## VII. Summary

In conclusion, we have performed a systematic investigation of the hole concentration p of single-layer BSLCO in two doping series, $Bi_2Sr_{2-x}La_xCuO_{6+\delta}$ and $Bi_{2-y}Pb_ySr_{2-x}La_xCuO_{6+\delta}$, by a detailed analysis of the oxygen ligand line at the Cu $L_3$-absorption edge. We have found the optimum hole doping level for maximum $T_c$ at p = 0.17 respective p = 0.16. The relationship between the ratio of $T_c/T_{c,max}$ and p shows a faster drop on both, the underdoped and the overdoped side of the phase diagram in variance with the so-called universal hole doping curve [15] of cuprate high-$T_c$ superconductors. This shows that the general trends are similar in all doped cuprates but in order to compare results on a high accuracy level the hole concentration has to be determined individually. More over effects like a drop of $T_c$ around a hole concentration of 1/8 will not be described by an universal curve.

Further on, we have determined the evolution of the occupied and unoccupied density of states at the Fermi energy with doping as derived from PES and XAS spectra series, respectively. In the XAS series the so-called prepeak of the oxygen K-edge has been analysed. Both, the occupied as well as the unoccupied density of states reveal the same dependence on hole doping, i.e. a continuous increase in the hole underdoped region and a constant density in the hole overdoped region. The results of our extensive study on BSLCO are in line with previous data of LSCO [2 – 4]. From all single layer cuprates studied so far one can conclude that the evolution into constant density of states for the hole overdoped $CuO_2$-plane is established. This, on the other hand, is a hint that the $CuO_2$-plane consists of two types of doped holes, from which the mobile ones determine the intensity of the prepeak of the O 1s absorption edge and the superconductivity of single-layer cuprates.

### Acknowledgments

We gratefully acknowledge assistance by the staff of HASYLAB and BESSY, namely Dr. P. Gürtler, Dr. K. Roßnagel, Dr. D. Batchelor and Dr. P. Bressler. We also thank D. Kaiser for the crystal growth. This work was supported by the BMBF, Project No. 05 KS1 KHA 0.